\documentclass[5p,twocolumn,number,sort&compress]{elsarticle}
\usepackage[utf8]{inputenc}
\usepackage[T1]{fontenc}
\DeclareUnicodeCharacter{2009}{\,}
\DeclareUnicodeCharacter{202F}{\,}
\DeclareUnicodeCharacter{2010}{-}
\DeclareUnicodeCharacter{2212}{\ensuremath{-}}
\usepackage{graphicx}
\usepackage{amsmath,amssymb,bm}
\usepackage{booktabs}
\usepackage{xcolor}
\usepackage{docmute}
\newcommand{\inmaindocument}{}
\usepackage[colorlinks,linkcolor=blue,citecolor=blue,urlcolor=blue]{hyperref}
\newcommand{\kL}{\kappa}
\journal{Computational Materials Science}
\begin{document}
\begin{frontmatter}
\title{Defect-limited thermal transport in AlN using pretrained machine-learning interatomic potentials}
\author[snu]{Minseok Moon\texorpdfstring{\fnref{equal}}{}}
\ead{njs03019@snu.ac.kr}
\author[snu]{Wonjun Choi\texorpdfstring{\fnref{equal}}{}}
\ead{wonjun.choi@snu.ac.kr}
\author[snu,riam,kias]{Seungwu Han\texorpdfstring{\corref{cor}}{}}
\ead{hansw@snu.ac.kr}
\author[inu]{Youngho Kang\texorpdfstring{\corref{cor}}{}}
\ead{youngho84@inu.ac.kr}
\address[snu]{Department of Materials Science and Engineering,
Seoul National University, Seoul, 08826, Republic of Korea}
\address[riam]{Research Institute of Advanced Materials,
Seoul National University, Seoul, 08826, Republic of Korea}
\address[kias]{Center for AI and Natural Sciences,
Korea Institute for Advanced Study, Seoul, 02455, Republic of Korea}
\address[inu]{Department of Materials Science and Engineering,
Incheon National University, Incheon 22012, Republic of Korea}
\fntext[equal]{These authors contributed equally to this work.}
\cortext[cor]{Corresponding authors.}
\begin{abstract}
Aluminum nitride (AlN) is an important thermal management material whose high lattice thermal conductivity is strongly suppressed by oxygen impurities. We investigate phonon scattering by oxygen-related defects using pretrained universal machine-learning interatomic potentials (MLIPs), molecular dynamics (MD), and phonon Boltzmann transport calculations. Several pretrained MLIPs are benchmarked against density functional theory for phonon dispersions and pristine thermal conductivity. To balance accuracy and computational speed, we use a fine-tuned version of the compact SevenNet-Nano model for MD simulations. Monte Carlo annealing supports the formation of bound $V_{\mathrm{Al}}(\mathrm{O_N})_3$ complexes, whose scattering differs from that of their isolated constituents. Defect scattering rates extracted from excess spectral energy density (SED) linewidths agree reasonably with harmonic $T$-matrix predictions at low oxygen contents, supporting the independent-scatterer approximation. Incorporating these rates into an iterative Boltzmann transport equation with Bose--Einstein statistics yields thermal conductivities comparable to experimental values and captures their observed decrease with oxygen content. At high concentrations, however, the scattering rate deviates from linear scaling with oxygen content, suggesting limitations of independent-defect and pristine-phonon descriptions. These results demonstrate an efficient approach using pretrained MLIPs to quantify defect-limited thermal transport and provide insights into impurity effects beyond the dilute-defect approximation.
\end{abstract}
\begin{keyword}
Aluminum nitride \sep Thermal conductivity \sep Point defects
\sep Spectral energy density \sep Boltzmann transport equation
\sep Machine-learning interatomic potentials

\end{keyword}
\end{frontmatter}

%======================================================================
% Introduction (formerly sections/01_intro.tex)
%======================================================================
\section{Introduction}\label{sec:intro}

Aluminum nitride (AlN) is a technologically important high-thermal-conductivity ceramic widely used in electronic packaging and thermal management, particularly as an electrically insulating substrate for high-power semiconductor devices~\cite{kurokawa1988, jackson1997}. Its combination of high thermal conductivity, electrical insulation, and a relatively low thermal expansion coefficient makes it attractive for efficient heat dissipation in advanced electronic devices~\cite{zhang2024aln}. Its room-temperature thermal conductivity $\kL$ is among the highest of nonmetallic crystals and comparable to that of metals~\cite{slack1973}. Reported room-temperature values reach up to 374~W\,m$^{-1}$\,K$^{-1}$~\cite{rounds2018apex}.

However, the thermal conductivity of AlN is highly sensitive to impurities and crystalline defects, which introduce additional phonon scattering and can substantially suppress heat transport~\cite{rounds2018,xu2019}. Oxygen-related defects are particularly important because oxygen incorporation during synthesis and processing is difficult to avoid~\cite{jackson1997,virkar1989,zhang2024aln}. Despite their substantial impact on thermal conductivity, the phonon scattering mechanisms associated with these defects remain incompletely understood. Previous theoretical studies have examined individual point defects, including Al vacancies ($V_{\mathrm{Al}}$) and substitutional oxygen on N sites ($\mathrm{O_N}$)~\cite{dou2025,li2025aln}. Moreover, oxygen impurities can associate with Al vacancies to form bound defect complexes~\cite{harris1990,youngman1990,mattila1997}, whose scattering behavior may differ from that of isolated constituent defects. The mode-resolved phonon scattering induced by these complexes and its resulting impact on the lattice thermal conductivity of AlN have yet to be systematically characterized.

The influence of point defects on phonon transport has been studied using first-principles approaches, primarily through Fermi's golden rule (FGR) and Green's-function-based $T$-matrix formulations~\cite{katcho2014,li2025aln}. Defect scattering calculated using FGR typically employs the lowest-order Born approximation, whereas the $T$-matrix formulation sums repeated scattering from the defect perturbation to all orders, allowing strong scattering and resonant effects to be described~\cite{katcho2014,katre2017}. The latter has also been applied to bound defect complexes in GaN and ThO$_2$~\cite{katre2018,malakkal2024}. However, conventional calculations based on a single-defect $T$ matrix generally assume dilute, independent scattering centers, yielding scattering rates that scale linearly with defect concentration~\cite{thebaud2020}. The validity of this approximation at higher concentrations, where structural perturbations may overlap and the vibrational spectrum may change appreciably, requires further investigation~\cite{thebaud2020}.

Recent advances in machine-learning interatomic potentials (MLIPs) have enabled efficient molecular dynamics (MD) simulations with accuracy approaching that of computationally demanding first-principles calculations~\cite{batzner2022,musaelian2023,park2024}. In particular, pretrained universal MLIPs such as MACE-MH-1~\cite{batatia2025crosslearning}, eSEN~\cite{fu2025esen}, and SevenNet-Omni~\cite{kim2026omni}, trained on extensive and diverse datasets of atomic structures, have broadened access to simulations of various materials without system-specific training from scratch. These developments facilitate investigations of phonon transport using MD-based methods such as spectral energy density (SED) analysis, which extracts phonon frequencies and linewidths directly from atomic trajectories~\cite{thomas2010,larkin2014}. Unlike conventional single-defect $T$-matrix calculations, MD simulations underlying SED analysis explicitly include multiple defects and their associated structural and vibrational modifications, without imposing independent defect scattering~\cite{sun2025,chen2024}. Comparisons of SED-derived scattering rates and the resulting thermal conductivities with $T$-matrix predictions, together with defect-induced changes in the SED spectra, can therefore provide insights into the validity and limitations of the dilute-defect approximation as defect concentration increases~\cite{zhou2024,chen2024}.

In this study, we investigate how oxygen-related defect complexes affect phonon scattering and lattice thermal conductivity in AlN using pretrained universal MLIPs. We benchmark their predictions of phonon dispersions and pristine thermal conductivity against density functional theory (DFT) and fine-tune a compact model to achieve accuracy comparable to its teacher model at substantially lower computational cost. For the energetically favored $V_{\mathrm{Al}}(\mathrm{O_N})_3$ complexes, whose formation is also supported by Monte Carlo annealing, we extract defect-induced scattering rates from MD simulations using SED analysis at several oxygen contents and compare them with harmonic $T$-matrix predictions. These rates are then incorporated into an iterative solution of the phonon Boltzmann transport equation (BTE)~\cite{omini1995,li2014} with Bose--Einstein phonon statistics, and the resulting dependence of thermal conductivity on oxygen content is compared with reported measurements~\cite{slack1973,rounds2018,kim2022}. Finally, we assess the validity of the dilute-defect approximation as oxygen content increases and discuss available approaches to modeling phonon transport beyond this approximation, together with their capabilities and limitations.

%======================================================================
% Methods (formerly sections/02_protocol.tex)
%======================================================================
% =====================================================================
% §II 프로토콜 정의 (기획문서 §2) — 시스템-불가지론적 서술이 생명:
% "AlN에서 1.280을 곱한다"가 아니라 "임의 계에서 이 절차로 실측한다".
% =====================================================================
\section{Methods}\label{sec:protocol}

\subsection{Lattice thermal conductivity from the BTE}
\label{sec:correction}

The lattice thermal conductivity is calculated by iteratively solving the linearized phonon  BTE~\cite{omini1995} using a modified ShengBTE package~\cite{li2014}. A phonon mode is denoted by $\lambda=(\mathbf{q},s)$, where $\mathbf{q}$ and $s$ are its wave vector and branch index, respectively. Under a small temperature gradient, the deviation of the phonon population from its equilibrium Bose--Einstein distribution $n_\lambda^0$ is expressed as
\begin{equation}
n_\lambda-n_\lambda^0
=-\frac{\partial n_\lambda^0}{\partial T}
\sum_\beta F_\lambda^\beta\nabla_\beta T.
\label{eq:population_deviation}
\end{equation}
The response $F_\lambda^\beta$ is obtained, for each Cartesian direction $\beta$, by solving
\begin{equation}
v_\lambda^\beta
=\sum_{\lambda'}\Omega_{\lambda\lambda'}F_{\lambda'}^\beta,
\label{eq:linearized_bte}
\end{equation}
where $v_\lambda^\beta$ is the $\beta$ component of the phonon group velocity and $\Omega_{\lambda\lambda'}$ is the linearized scattering operator. Separating its diagonal and mode-coupling terms as $\Omega_{\lambda\lambda'}=\delta_{\lambda\lambda'}/\tau_\lambda-P_{\lambda\lambda'}$ gives the iterative update
\begin{equation}
F_\lambda^{\beta,(i+1)}
=\tau_\lambda\left[
v_\lambda^\beta+
\sum_{\lambda'}P_{\lambda\lambda'}
F_{\lambda'}^{\beta,(i)}
\right],
\label{eq:bte_iteration}
\end{equation}
with the relaxation-time approximation (RTA) result $F_\lambda^{\beta,(0)}=\tau_\lambda v_\lambda^\beta$ as the initial guess. Here, $\tau_\lambda^{-1}$ denotes the total mode-specific scattering rate, including phonon--phonon and defect scattering contributions, whereas $P_{\lambda\lambda'}$ represents the mode-coupling terms associated with phonon--phonon scattering in the iterative BTE solution. After convergence, the lattice thermal conductivity tensor is evaluated as
\begin{equation}
\kL_{\alpha\beta}
=\frac{1}{V}\sum_\lambda
c_\lambda v_\lambda^\alpha F_\lambda^\beta,
\label{eq:modesum}
\end{equation}
where $V$ is the crystal volume and $c_\lambda$ is the modal heat capacity. For wurtzite AlN, the thermal conductivity tensor is diagonal with two independent components, corresponding to transport in the basal plane ($\kL_{xx}=\kL_{yy}$) and along the $c$ axis ($\kL_{zz}$). The BTE is solved at 300~K on a $26\times26\times24$ $\mathbf{q}$-point mesh, which provides converged thermal conductivities (Fig.~\ref{smfig:qmesh})~\cite{dou2025}. The iterative solution yields a thermal conductivity approximately 10\% higher than the RTA result (Table~\ref{smtab:solvercomparison}). This increase reflects the coupling between the nonequilibrium populations of different phonon modes, which allows momentum-conserving normal processes to redistribute phonons without relaxing the heat current as effectively as assumed in the RTA.

\subsection{Scattering rate}
\label{sec:scattering rate}

We consider two scattering mechanisms: phonon--phonon and phonon--impurity scattering. The phonon--phonon scattering rate, $\Gamma_{\mathrm{ph},\lambda}$, is calculated from three-phonon anharmonic interactions using FGR. The second-order force constants are obtained by the finite-displacement method using phonopy~\cite{togo2015b} in a $5\times5\times3$ supercell. The third-order force constants are calculated using the \texttt{thirdorder.py} script of ShengBTE~\cite{li2012thirdorder,li2014} in a $4\times4\times3$ supercell. We retain third-order interactions for atomic triplets whose three pairwise interatomic distances are all within 5.0~\AA{} and impose the acoustic sum rule during force-constant reconstruction. A displacement amplitude of 0.03~\AA{} is used to calculate both sets of force constants. The energy-conserving delta functions in the three-phonon scattering rates are approximated using the adaptive Gaussian smearing scheme implemented in ShengBTE~\cite{li2012}. 

The defect scattering rate $\Gamma_{\mathrm{def},\lambda}$ is obtained in two independent ways, from the harmonic scattering $T$-matrix and from the SED in MD simulations. In the $T$-matrix approach, the defect is represented by a static harmonic perturbation, $\mathbf{V}(\omega)=\mathbf{V}_{\mathrm{K}}-\omega^{2}\mathbf{S}$, where $\mathbf{V}_{\mathrm{K}}$ is the difference in mass-weighted force constants between the defective and pristine supercells, and $\mathbf{S}$ is a diagonal matrix describing the relative mass change~\cite{malakkal2024,dou2025}. The force-constant difference is obtained from finite-displacement calculations with a displacement
amplitude of 0.01~\AA{} in a $6\times6\times4$ supercell (576 atomic sites) containing one defect. No spatial cutoff is applied to the force-constant difference within the supercell. Multiple scattering by the defect is included to all orders through the $T$ matrix,
\begin{equation}
\mathbf{T}(\omega)
=\left[\mathbf{I}-\mathbf{V}(\omega)\mathbf{G}_{0}(\omega)\right]^{-1}
\mathbf{V}(\omega),
\label{eq:tmat}
\end{equation}
where $\mathbf{G}_{0}(\omega)$ is the retarded Green function of the pristine crystal. It is evaluated on a $64\times64\times35$ $\mathbf{q}$-point mesh with a broadening of 0.10~THz. The defect-induced scattering rate for a given number of defects per unit cell ($c_{\mathrm{def}}$) is obtained as~\cite{li2025aln}
\begin{equation}
\Gamma_{\mathrm{def},\lambda}
=-\frac{c_{\mathrm{def}}}{\omega_{\lambda}}\,
\mathrm{Im}\,
\langle\lambda|\mathbf{T}(\omega_{\lambda})|\lambda\rangle,
\label{eq:tmrate}
\end{equation}
where $|\lambda\rangle$ is a pristine-crystal phonon eigenvector normalized to one primitive cell and $\omega_{\lambda}=2\pi\nu_{\lambda}$ is its angular frequency.

In the SED approach, the defect scattering rate is estimated from the excess spectral linewidth of the defective supercell relative to the pristine supercell in MD simulations. The SED, which represents the phonon spectral lineshape, is calculated from mass-weighted plane-wave projections of the atomic velocities~\cite{thomas2010,larkin2014}:
\begin{equation}
\Phi(\mathbf{q},\nu)\propto
\sum_{b,\alpha}
\left|
\int_{0}^{t_{0}} W(t)
\sum_{j\in b}\sqrt{m_{j}}\,v_{j}^{\alpha}(t)\,
e^{\mathrm{i}\mathbf{q}\cdot\mathbf{r}_{j}^{0}
-2\pi\mathrm{i}\nu t}\,dt
\right|^{2},
\label{eq:sed}
\end{equation}
where $m_j$, $v_j^\alpha(t)$, and $\mathbf{r}_j^0$ are the mass, Cartesian velocity component, and ideal lattice position of atom $j$ on sublattice $b$, respectively. The trajectory length is $t_0=4$~ns, and $W(t)=\sin^2(\pi t/t_0)$ is the Hann window. This plane-wave projection provides the SED on the wave-vector grid commensurate with the simulation supercell.

\begin{figure*}[htbp]
\makebox[\textwidth][c]{\includegraphics{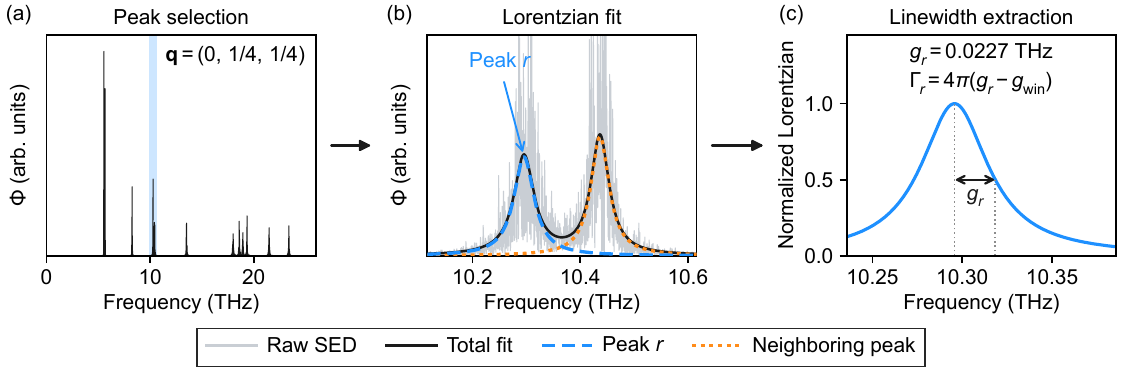}}
\caption{Extraction of phonon scattering rates from the SED of pristine AlN: (a) SED intensity $\Phi(\mathbf{q},\nu)$ at $\mathbf{q}=(0,1/4,1/4)$, with the selected fitting interval shaded; (b) simultaneous Lorentzian fitting of overlapping peaks, including the target peak $r$; and (c) the Lorentzian component of peak $r$, normalized to unit height. The double-headed arrow in (c) indicates the half width at half maximum, $g_r=0.0227$~THz, measured from the peak center to a half-maximum point. The scattering rate is calculated as $\Gamma_r=4\pi(g_r-g_{\mathrm{win}})$ using the approximate window-broadening correction in Eq.~\eqref{eq:winrate}.}
\label{fig:sedextraction}
\end{figure*}

Several peaks corresponding to different phonon branches are observed in $\Phi(\mathbf{q},\nu)$, as shown in Fig.~\ref{fig:sedextraction}, which illustrates the scattering-rate extraction procedure. Each SED peak $r$ is fitted with a Lorentzian function and a constant background within a frequency interval of $\pm1.0$~THz around its position. Peaks with overlapping fitting intervals are fitted simultaneously. The fitted peaks are assigned to the harmonic phonon modes of the pristine crystal at the same wave vector by matching their frequency ordering. For each resolved peak, the fitted half width at half maximum $g_r$ (in THz) is converted to a scattering rate (in ps$^{-1}$) after subtracting the Hann-window broadening:
\begin{equation}
\Gamma_r=4\pi(g_r-g_{\mathrm{win}}),
\qquad g_{\mathrm{win}}=\frac{0.75}{t_0}.
\label{eq:winrate}
\end{equation} 

Assuming that the phonon--phonon and defect scattering rates are additive (Matthiessen's rule), the defect scattering rate associated with SED peak $r$ is obtained by subtracting the pristine rate from the defective rate:
\begin{equation}
\Gamma_{\mathrm{def},r}=\bar{\Gamma}_{r}^{\,\mathrm{defect}}-\bar{\Gamma}_{r}^{\,\mathrm{pristine}},
\label{eq:gdef}
\end{equation}
where the overbars denote averages over three independent MD runs for each supercell. 

For each approach, the defect scattering rates are combined with the anharmonic phonon scattering rates according to Matthiessen's rule and incorporated into the iterative BTE solution to evaluate the lattice thermal conductivity. For the $T$-matrix approach, Eq.~\eqref{eq:tmrate} is evaluated directly on the $26\times26\times24$ $\mathbf{q}$-point mesh used to solve the BTE, using pristine-crystal phonon eigenvectors. In contrast, the SED-derived defect scattering rates are obtained from MD simulations in a $12\times12\times8$ supercell and are therefore available only at wave vectors commensurate with that supercell. To assign rates to modes on the denser BTE mesh, we use the weighted median of the sampled SED rates. A sampled SED peak $r$, with wave vector $\mathbf{q}_r$ and frequency $\nu_r$, is assigned the weight
\begin{equation}
w_r=\exp\!\left[-\frac{1}{2}\left(\frac{\Delta q_r^2}{h_q^2}+\frac{(\nu_\lambda-\nu_r)^2}{h_\nu^2}\right)\right],
\label{eq:kernel}
\end{equation}
where $\Delta q_r$ is the distance between the wave vector of phonon mode $\lambda$ and the nearest symmetry-equivalent image of $\mathbf{q}_r$. The weights thus account for proximity in both wave vector and frequency. The kernel widths $h_q$ and $h_\nu$ are selected by leave-one-wave-vector-out cross-validation.

At low frequencies, the available wave-vector sampling may be insufficient to resolve the defect scattering rates reliably. Since scattering by static, localized defects is elastic and follows the Rayleigh law in the long-wavelength limit~\cite{klemens1955,ratsifaritana1987,tamura1983}, we extrapolate the rates below 2.5~THz from their values at 2.5~THz using $\Gamma_{\mathrm{def}}\propto\nu^4$. The same low-frequency extrapolation is applied to the $T$-matrix rates.

\subsection{MLIP calculations}
\label{sec:potentials}

We examine five MLIPs with a message-passing equivariant architecture, four pretrained and one fine-tuned: three of the SevenNet family~\cite{park2024} and two of the MACE family~\cite{batatia2022}. Of the SevenNet family, the first, omni-mpa, is the mpa channel of the pretrained multitask universal model SevenNet-Omni~\cite{kim2026omni,kim2025}. This channel corresponds to the Perdew--Burke--Ernzerhof (PBE) exchange--correlation functional used in our DFT calculations (Section~\ref{sec:dft}). The second, nano, is the 5.0~\AA{} cutoff variant of SevenNet-Nano~\cite{oh2026nano}, a compact model distilled from the mpa channel of SevenNet-Omni. The third, ft-nano, is obtained by fine-tuning nano on configurations labeled by single-point evaluations with omni-mpa, bringing its predictions closer to those of omni-mpa for the structures considered here. The fine-tuning parameters and protocol are provided in Table~\ref{smtab:ftproto}, and the improvement in force predictions is shown in the parity plots in Fig.~\ref{smfig:parity}. Among the MACE models, MACE-MH-1~\cite{batatia2025crosslearning} is a multi-head model, for which we use the \texttt{omat\_pbe} head corresponding to the PBE/PBE+$U$ level of theory. MACE-MPA-0~\cite{batatia2024mp0} is a lighter single-head model trained on crystalline structures labeled using PBE/PBE+$U$ calculations.

Atomic forces used to calculate the force constants and perform structural relaxations are obtained with the Atomic Simulation Environment (ASE)~\cite{larsen2017}. MD simulations are performed using LAMMPS~\cite{thompson2022}. For SevenNet models, we employ FlashTP tensor-product kernels~\cite{lee2025flashtp} to accelerate  calculations. All MD simulations employ a time step of 1~fs.
The simulation cells are equilibrated at 300~K for 60~ps in the NVT ensemble using a Nos\'e--Hoover
thermostat~\cite{nose1984,hoover1985}. MD trajectories to extract SED are then generated for 4~ns in the NVE ensemble at the same fixed cell dimensions. The SED calculations use $12\times12\times8$ supercells of the primitive cell (4608 atomic sites). The defective supercells contain 8 and 16 $V_{\mathrm{Al}}(\mathrm{O_N})_3$ complexes, corresponding to oxygen contents of 0.407 and 0.816~wt\%. Atomic velocities are recorded every 10~fs for the SED analysis.

\subsection{DFT calculations}
\label{sec:dft}

All DFT calculations are performed using the Vienna Ab initio Simulation Package
(VASP)~\cite{kresse1995,kresse1996b,kresse1996,kresse1999} with the projector-augmented-wave method~\cite{blochl1994} and the PBE exchange-correlation
%Perdew--Burke--Ernzerhof exchange-correlation
functional~\cite{pbe1996}. The plane-wave cutoff energy is 520~eV and the Brillouin zone of the four-atom wurtzite cell is sampled with a $\Gamma$-centered $9\times9\times6$ $k$-point mesh. The electronic loop is converged to $10^{-8}$~eV. Both the ions and the lattice vectors are
relaxed until the residual forces fall below $10^{-4}$~eV/\AA. The $k$-point mesh for each force-constant supercell is adjusted according to the supercell dimensions while maintaining a comparable reciprocal-space sampling density.

%======================================================================
% Results: potentials, spectra, T matrix (formerly sections/03_validation.tex)
%======================================================================
\section{Results and discussion}
\label{sec:validation}

\subsection{MLIP validation}
\label{sec:mlipval}
\label{sec:reference}

We compare the lattice parameters calculated using DFT and the five MLIPs (Table~\ref{tab:reference}). Our DFT lattice parameters agree with the previous PBE calculation~\cite{dou2025} within 0.2\%. We find that omni-mpa and MACE-MH-1 excellently reproduce the DFT values within 0.1\%. The deviations of nano and MACE-MPA-0 are slightly larger, although their lattice parameters remain close to the DFT values. Fine-tuning of nano improves the agreement, with ft-nano yielding lattice parameters closer to those predicted by omni-mpa. 

\begin{table}[htbp]
\caption{Lattice parameters and lattice thermal conductivities of pristine AlN. Thermal conductivities are calculated at 300~K using the iterative BTE solution with anharmonic three-phonon scattering and are reported in W\,m$^{-1}$\,K$^{-1}$. Values in parentheses are the DFT results obtained within the RTA.}
\label{tab:reference}
\small
\begin{tabular*}{\columnwidth}{@{\extracolsep{\fill}}lcccc@{}}
\toprule
Model & $a$ (\AA) & $c$ (\AA) & $\kL_{xx}$ & $\kL_{zz}$ \\
\midrule
DFT & 3.129 & 5.016 & \begin{tabular}{@{}c@{}}271.0\\(247.7)\end{tabular} & \begin{tabular}{@{}c@{}}261.2\\(236.5)\end{tabular} \\
omni-mpa  & 3.128 & 5.013 & 259.1 & 231.2 \\
nano      & 3.141 & 5.029 & 232.5 & 233.6 \\
ft-nano   & 3.128 & 5.012 & 259.5 & 230.7 \\
MACE-MH-1 & 3.128 & 5.016 & 257.0 & 245.3 \\
MACE-MPA-0 & 3.133 & 5.026 & 248.2 & 215.8 \\
\midrule
DFT & 3.132$^{a}$ & 5.024$^{a}$ & 254$^{a}$ & 222$^{a}$ \\
Expt.     & 3.113$^{b}$ & 4.982$^{b}$ & \multicolumn{2}{c}{216--374$^{c}$} \\
\bottomrule
\end{tabular*}
\par\smallskip
{\footnotesize $^{a}$From Ref.~\cite{dou2025}; conductivities are DFT-RTA results. $^{b}$Experimental lattice parameters from Ref.~\cite{slack1973}. $^{c}$Range of reported room-temperature values at oxygen contents below $10^{-2}$~wt\% (Fig.~\ref{fig:kappaliterature}); directional components were not resolved.\par}
\end{table}

\begin{figure*}[htbp]
\makebox[\textwidth][c]{\includegraphics{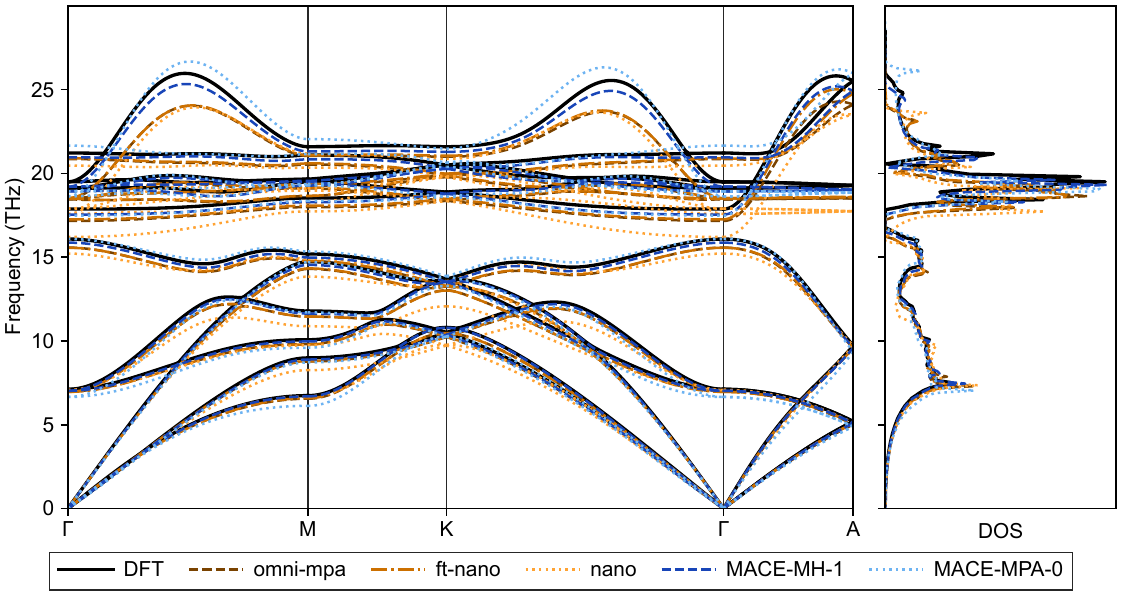}}
\caption{Harmonic phonon dispersions and densities of states of AlN calculated using DFT and MLIPs.}
\label{fig:dispval}
\end{figure*}

Fig.~\ref{fig:dispval} compares the harmonic phonon dispersions predicted by the MLIPs with those obtained from DFT. Omni-mpa and MACE-MH-1 closely reproduce the acoustic branches, which play a critical role in heat transport, along the plotted high-symmetry path, underestimating their frequencies by less than 2.5\%. Both models show larger underestimations for the optical branches, with more pronounced deviations for omni-mpa. Nano and MACE-MPA-0 underestimate the acoustic phonon frequencies more substantially than omni-mpa and MACE-MH-1. Nano also underestimates the optical phonon frequencies, similar to omni-mpa, whereas MACE-MPA-0 overestimates them relative to DFT. The fine-tuned model ft-nano yields phonon dispersions close to those obtained with omni-mpa.

Consistent with their somewhat softer acoustic modes relative to DFT, the MLIPs predict smaller group velocities and larger phonon scattering rates (Fig.~\ref{smfig:ratevg}), resulting in lower thermal conductivities for pristine AlN (Table~\ref{tab:reference}). Nevertheless, omni-mpa and MACE-MH-1 reproduce the DFT thermal conductivity to within about 10\%, whereas nano and MACE-MPA-0 show larger deviations. Fine-tuning improves the thermal conductivity predicted by nano, bringing the accuracy of ft-nano to a level comparable to that of omni-mpa. Our DFT thermal conductivities calculated within the RTA, given in parentheses in Table~\ref{tab:reference}, are consistent with previously reported calculations.

Extracting impurity scattering rates from SED linewidths requires long MD trajectories and large defective supercells, making an MLIP that balances accuracy and computational cost essential. Although omni-mpa and MACE-MH-1 provide relatively accurate predictions, they are computationally demanding. Nano runs approximately an order of magnitude faster than omni-mpa~\cite{oh2026nano}, but with lower accuracy. Ft-nano provides accuracy comparable to that of omni-mpa for the properties assessed above while retaining the computational efficiency of nano and is therefore used for the calculations presented below.

\begin{figure}[htbp]
\makebox[\columnwidth][c]{\includegraphics{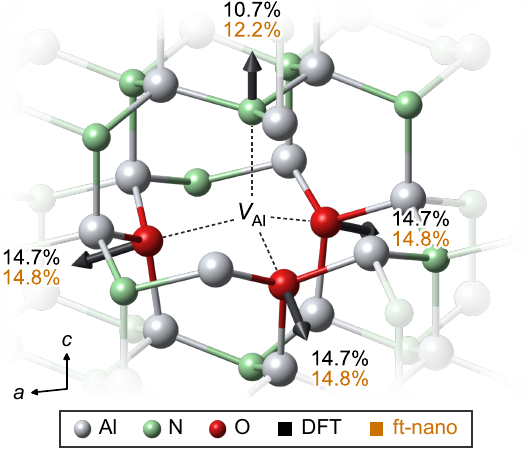}}
\caption{Relaxation of the four nearest neighbors of $V_{\mathrm{Al}}$ in the $V_{\mathrm{Al}}(\mathrm{O_N})_3$ complex. Arrows indicate the relaxation directions, which are common to DFT and ft-nano. Numbers give $(d-d_0)/d_0$ from DFT (black) and ft-nano (orange), where $d$ is the distance from the vacant site to each neighbor (dashed lines) and $d_0$ is the corresponding Al--N bond length in bulk AlN.}
\label{fig:defgeom}
\end{figure}

We consider complexes of $V_{\mathrm{Al}}$ and $\mathrm{O_N}$ as phonon scattering centers. Earlier defect-cluster models and first-principles studies support the energetic preference for vacancy--oxygen association~\cite{harris1990,youngman1990,mattila1997}. Up to three $\mathrm{O_N}$ defects can bind to an Al vacancy, forming $V_{\mathrm{Al}}(\mathrm{O_N})_n$ complexes with $n=1$, 2, or 3. To determine which complexes are preferentially formed at experimentally relevant oxygen contents, we perform Monte Carlo annealing at at 0.407~wt\% oxygen. Initially dispersed $\mathrm{O_N}$ impurities associate with Al vacancies to form $V_{\mathrm{Al}}(\mathrm{O_N})_3$ complexes (Fig.~\ref{fig:mcanneal}), supporting their thermodynamic preference under the simulated conditions. Ft-nano predicts a binding energy of 4.33~eV, compared with the DFT value of 4.81~eV, where the binding energy is defined as the energy required to separate the complex into isolated constituent point defects. Ft-nano also reproduces the DFT-relaxed local geometry of the complex, in which neighboring anions relax slightly outward (Fig.~\ref{fig:defgeom}). In what follows, we assume that all oxygen atoms incorporated into AlN form $V_{\mathrm{Al}}(\mathrm{O_N})_3$ defect complexes. 

\subsection{Phonon spectrum and scattering rate from the SED approach}
\label{sec:sedanalysis}
\begin{figure*}[htbp]
\makebox[\textwidth][c]{\includegraphics{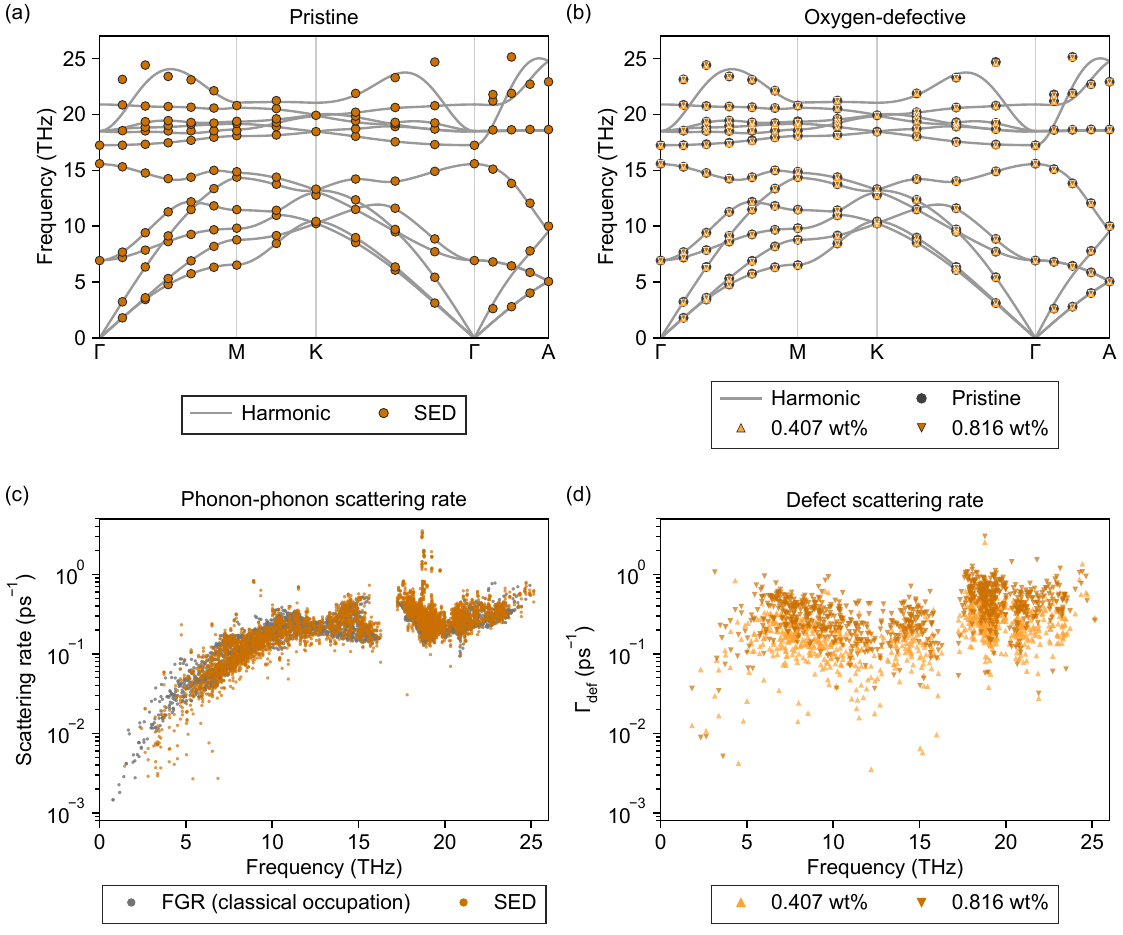}}
\caption{Phonon spectra and scattering rates of pristine (left) and oxygen-containing (right) AlN calculated using ft-nano. (a, b) SED peak positions for (a) the pristine supercell and (b) defective supercells with 0.407 and 0.816~wt\% oxygen, compared with the harmonic phonon dispersion of pristine AlN. (c) SED-derived and Fermi's golden rule (FGR) phonon--phonon scattering rates of pristine AlN, with the FGR rates evaluated using classical phonon occupations. (d) Defect scattering rates, $\Gamma_{\mathrm{def}}$, extracted from the excess SED linewidths at 0.407 and 0.816~wt\% oxygen.}
\label{fig:sedvalidate}\label{fig:defectrates}   % Figures 4 and 5 combined (10-04); old Figure 5 environment commented out below
\end{figure*}

We first assess whether the peak positions and linewidths extracted from the SED obtained by MLIP-MD simulations accurately capture the phonon spectrum and phonon--phonon scattering rates of pristine AlN. Fig.~\ref{fig:sedvalidate}(a) compares the SED peak positions (orange dots) with the harmonic phonon dispersion (gray curves). The two agree well overall, although some discrepancies are observed for high-frequency optical modes. Fig.~\ref{fig:sedvalidate}(c) compares the SED-derived scattering rates with phonon--phonon scattering rates calculated using FGR. For this comparison, the FGR rates are evaluated using classical phonon occupations rather than Bose--Einstein occupations, consistent with the classical MD simulations. The two sets of rates show good overall agreement in both magnitude and frequency dependence, supporting the use of SED linewidths to estimate phonon--phonon scattering rates in pristine AlN.

To evaluate the defect scattering rate from the SED approach, we also examine the SED peak positions of the defective systems. Fig.~\ref{fig:defectrates}(b) compares the peak positions at two oxygen concentrations, 0.407 and 0.816~wt\%, with the pristine AlN phonon dispersion. At both oxygen contents, the SED peak positions largely follow the pristine phonon dispersion, supporting the use of pristine-crystal phonon modes as a reference for describing transport at the defect concentrations considered. As shown in Fig.~\ref{fig:defectrates}(d), the defect-induced scattering rates, extracted independently at each oxygen content, increase approximately in proportion to the oxygen content, consistent with treating the bound complexes as independent scattering centers over the concentrations examined.

\begin{figure}
\centering
\includegraphics[width=90mm]{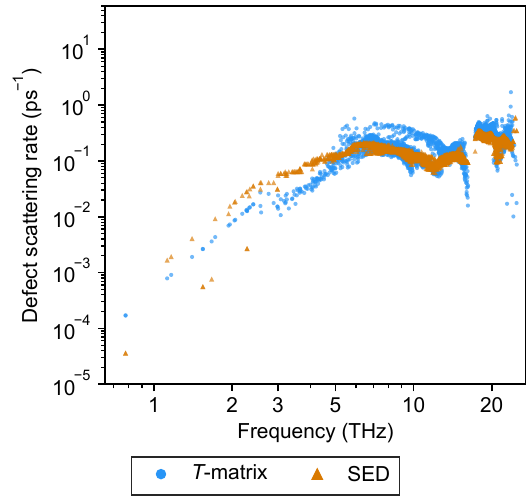}
\caption{Defect scattering rates at 0.407~wt\% oxygen from SED interpolated onto the BTE mesh and harmonic $T$-matrix calculations performed directly at the BTE wave vectors. Both use $\nu^4$ extrapolation below 2.5~THz.}
\label{fig:ratevalid}
\end{figure}

The defect-induced scattering rates extracted from the SED analysis are compared with the harmonic $T$-matrix rates of the relaxed $V_{\mathrm{Al}}(\mathrm{O_N})_3$ complex (0.407~wt\%). Fig.~\ref{fig:ratevalid} shows that the two sets of rates are comparable with each other over the entire frequency range. In the SED approach, multiple defect complexes are distributed throughout a large supercell at the specified defect content, whereas the $T$-matrix approach scales the single-complex elastic scattering rate by the corresponding number of complexes per primitive cell. Agreement between the two approaches therefore supports treating the complexes as independent elastic scattering centers over the concentration range considered and is consistent with the additive rate decomposition assumed in Matthiessen's rule [Eq.~\eqref{eq:gdef}].

%======================================================================
%======================================================================
\subsection{Thermal conductivity versus oxygen content}
\label{sec:conc}

\begin{figure*}[htbp]
\makebox[\textwidth][c]{\includegraphics{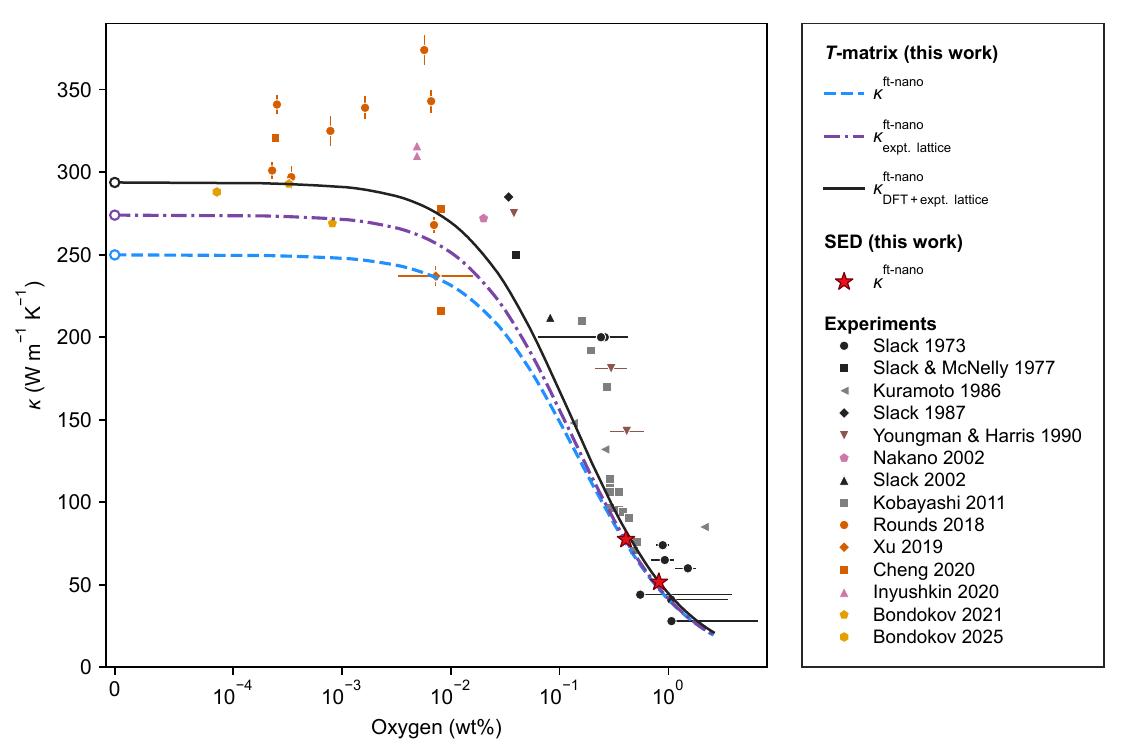}}
\caption{Thermal conductivity of AlN as a function of oxygen concentration at 300~K. Calculated values are orientational averages, $(2\kL_{xx}+\kL_{zz})/3$, from the iterative BTE. Lines show results with $T$-matrix defect rates obtained with ft-nano at the relaxed and experimental lattice parameters, and the latter further multiplied, for each direction, by the ratio of the DFT to omni-mpa pristine conductivities of Table~\ref{tab:reference}. Stars show results with SED defect rates obtained with ft-nano at its relaxed lattice parameters, and open circles at zero oxygen denote calculated pristine AlN. Experimental data are adopted from Refs.~\cite{slack1973,slack1977,kuramoto1986,slack1987,youngman1990,nakano2002,slack2002,kobayashi2011,rounds2018,rounds2018apex,xu2019,cheng2020,inyushkin2020,bondokov2021,bondokov2025}. Horizontal bars indicate reported oxygen ranges and vertical bars reported uncertainties.}
\label{fig:kappaliterature}
\end{figure*}

Fig.~\ref{fig:kappaliterature} compares the calculated lattice thermal conductivity as a function of oxygen concentration with reported measurements. As discussed above, we assume that all incorporated oxygen forms $V_{\mathrm{Al}}(\mathrm{O_N})_3$ complexes. Since most experimental studies did not report direction-resolved thermal conductivities, we focus on the orientational average, $(2\kL_{xx}+\kL_{zz})/3$, calculated using ft-nano. The three curves are iterative BTE results in which the $T$-matrix defect scattering rates are scaled in proportion to the oxygen content. In addition to the results at the relaxed ft-nano lattice parameters ($\kappa^{\mathrm{ft\mbox{-}nano}}$), we present two estimates with empirical corrections. The first accounts for the sensitivity of thermal conductivity to lattice parameters. PBE and the MLIPs trained on PBE data overestimate the lattice parameters, resulting in phonon softening, particularly for the optical modes (Fig.~\ref{smfig:displattice}). The lower optical phonon frequencies contribute to an approximately 11\% increase in the phonon--phonon scattering rates of the heat-carrying acoustic modes, thereby reducing the lattice thermal conductivity (Fig.~\ref{smfig:ratevglattice}). Using the experimental lattice parameters increases the calculated conductivity by 10\% ($\kappa_{\mathrm{expt.\ lattice}}^{\mathrm{ft\mbox{-}nano}}$; Table~\ref{smtab:latticebte}). The second correction accounts for the remaining difference between the DFT and MLIP predictions. In addition to the lattice-parameter discrepancy, residual errors in the phonon spectra predicted by omni-mpa and, consequently, ft-nano contribute to their underestimation of the lattice thermal conductivity relative to DFT. We therefore further rescale the conductivity obtained with the experimental lattice parameters by the ratio of the DFT to omni-mpa pristine conductivities in Table~\ref{tab:reference} ($\kappa_{\mathrm{DFT+expt.\ lattice}}^{\mathrm{ft\mbox{-}nano}}$).

The reported thermal conductivity of AlN with low oxygen contents ($<10^{-2}$~wt\%) ranges from 216 to 374~W\,m$^{-1}$\,K$^{-1}$. This spread may reflect differences in sample quality and measurement conditions, as well as measurement uncertainty. Our corrected estimate, $\kappa_{\mathrm{DFT+expt.\ lattice}}^{\mathrm{ft\mbox{-}nano}}$, is approximately 294~W\,m$^{-1}$\,K$^{-1}$, in reasonable agreement with the mean of the reported experimental values. The calculated conductivity begins to decrease appreciably above an oxygen content of $10^{-2}$~wt\%, indicating that impurity scattering becomes comparable in importance to anharmonic phonon scattering. Above approximately 0.2~wt\%, it becomes the dominant mechanism limiting the thermal conductivity. At the two concentrations examined, the SED approach yields orientationally averaged conductivities of 77.5~W\,m$^{-1}$\,K$^{-1}$ at 0.407~wt\% and 51.6~W\,m$^{-1}$\,K$^{-1}$ at 0.816~wt\%, within 10\% of the $T$-matrix values obtained with the same potential and ft-nano lattice parameters (75.0 and 47.4~W\,m$^{-1}$\,K$^{-1}$, respectively). Our results highlight the importance of reducing the oxygen content in AlN ceramics to achieve high thermal conductivity. 

\subsection{Outlook}
\label{sec:outlook}

\begin{figure*}[htbp]
\centering
\makebox[\textwidth][c]{\includegraphics{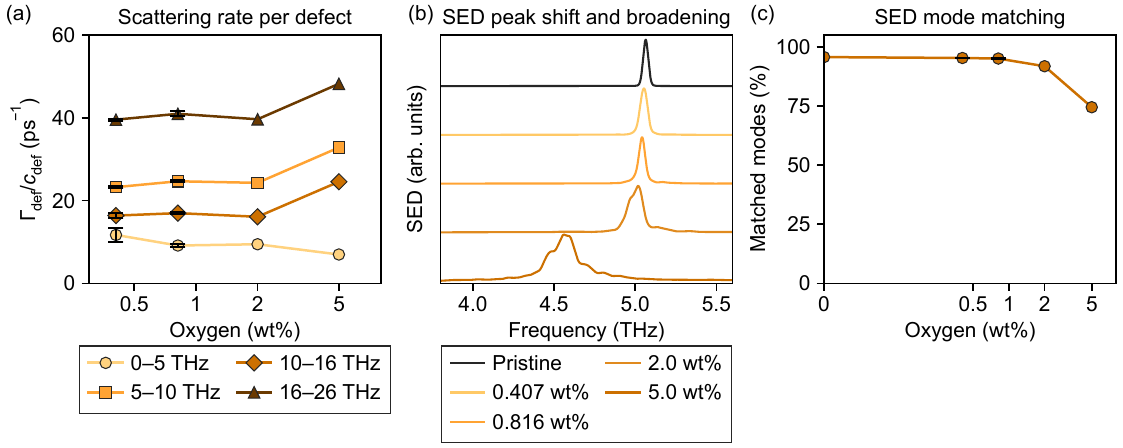}}
\caption{SED analysis at high oxygen contents. (a) SED excess scattering rate per defect complex, $\Gamma_{\mathrm{def}}/c_{\mathrm{def}}$, given as the median over the matched SED peaks in four frequency ranges. (b) SED intensity at the A point, $\mathbf{q}=(0,0,1/2)$, near 5~THz for pristine AlN and oxygen contents from 0.407 to 5~wt\%, normalized and offset vertically. (c) Fraction of the pristine harmonic phonon modes at the 95 SED wave vectors that are matched by a fitted SED peak. Symbols in (a) and (c) are averages over three independent MD runs at 0.407 and 0.816~wt\%  with bars indicating the standard error of the mean; the values at 2 and 5~wt\% and the spectra in (b) are from a single run.}
\label{fig:highO}
\end{figure*}

In the present work, we considered the $V_{\mathrm{Al}}(\mathrm{O_N})_3$ defect complex. Its phonon scattering rates differ from those of the isolated constituent defects and from the sum of their individual rates (Fig.~\ref{smfig:cvp}), indicating that the complex must be treated as a distinct scattering center. SED analysis of AlN with oxygen contents up to 5~wt\% reveals a deviation from linear concentration scaling between 2 and 5~wt\% oxygen (Fig.~\ref{fig:highO}(a)): the concentration-normalized rate, $\Gamma_{\mathrm{def}}/c_{\mathrm{def}}$, remains nearly constant up to 2~wt\% but changes substantially at 5~wt\%. This deviation is accompanied by pronounced frequency shifts and broadening of the SED peaks (Fig.~\ref{fig:highO}(b)), suggesting that a description based on pristine phonon modes becomes increasingly inadequate with increasing defect concentration. Together, these observations suggest a breakdown of the independent-scatterer approximation used in conventional single-defect $T$-matrix calculations and highlight the growing importance of disorder-induced changes in the vibrational modes. Although SED spectra can still be evaluated for highly disordered crystalline systems, extracting well-defined phonon lifetimes becomes increasingly challenging as mode mixing and spectral broadening obscure individual phonon peaks. This difficulty is reflected in the decreasing fraction of pristine phonon modes that can be matched to SED peaks (Fig.~\ref{fig:highO}(c)).

For such highly defective crystalline systems, alternative approaches that do not require assigning scattering rates to individual pristine phonon modes, such as Green--Kubo MD and lattice-dynamical approaches based on the quasi-harmonic Green--Kubo (QHGK) or Wigner formulations, may be more appropriate~\cite{isaeva2019,simoncelli2019,simoncelli2022}. However, classical Green--Kubo MD does not account for Bose--Einstein phonon statistics, which remain important for AlN even at room temperature. Indeed, our BTE calculations for pristine AlN based on phonon--phonon scattering show a difference of approximately 25\% in thermal conductivity between classical and Bose--Einstein phonon occupations under otherwise identical conditions (Fig.~\ref{smfig:classical}). Although QHGK and Wigner formulations can incorporate quantum phonon statistics and intermode transport contributions, their application to large disordered supercells may be computationally demanding, particularly when anharmonic phonon linewidths must also be evaluated. These limitations highlight the need for more computationally efficient approaches that can reliably describe thermal transport in strongly disordered crystalline systems, such as highly doped AlN, while retaining quantum phonon statistics.

%======================================================================
% Conclusions (formerly sections/05_summary.tex)
%======================================================================
\section{Conclusions}\label{sec:summary}
In this study, we investigated oxygen-related defect scattering and lattice thermal conductivity in AlN using pretrained universal MLIPs. Benchmarking against DFT identified accurate models, and fine-tuning SevenNet-Nano recovered accuracy comparable to omni-mpa while retaining the computational efficiency required for large-scale MD simulations. Monte Carlo annealing supports the formation of bound $V_{\mathrm{Al}}(\mathrm{O_N})_3$ complexes. Their scattering differs from the summed contributions of isolated constituents, demonstrating the importance of explicitly accounting for defect association. At low oxygen contents, $T$-matrix scattering rates agree reasonably with SED-derived rates, supporting the treatment of complexes as independent elastic scattering centers. The calculated thermal conductivity also captures the oxygen-content dependence observed experimentally. At higher oxygen contents, nonlinear scattering-rate scaling, spectral shifts, and broadening suggest limitations of independent-scatterer and pristine-phonon descriptions. These changes also hinder SED-based lifetime extraction, highlighting the challenges of extending dilute-defect transport calculations to strongly disordered systems. The alternative approaches discussed offer ways to address this regime, each with its own limitations. Overall, our results highlight the importance of reducing oxygen content to improve the thermal conductivity of AlN, establish an efficient MLIP-based route for predicting defect-limited thermal transport, and clarify the physical and methodological challenges beyond the dilute-defect approximation.

\section*{Acknowledgements}
This research was supported by the Nano \& Material Technology
Development Programs through the National Research Foundation of
Korea (NRF) funded by the Ministry of Science and ICT
(RS-2024-00407995 and RS-2026-25542918).

\bibliographystyle{elsarticle-num}
\bibliography{refs}

% Supplemental Material (combined PDF, also used for arXiv).
\clearpage

\onecolumn
 % SM 은 단일 컬럼. 이후 문서 끝까지 유지된다.
\setcounter{figure}{0}
\setcounter{table}{0}
\setcounter{section}{0}
\setcounter{equation}{0}
\renewcommand{\thefigure}{S\arabic{figure}}
\renewcommand{\thetable}{S\arabic{table}}
\renewcommand{\thesection}{S\arabic{section}}
\renewcommand{\theequation}{S\arabic{equation}}
\renewcommand{\theHfigure}{supp.\arabic{figure}}
\renewcommand{\theHtable}{supp.\arabic{table}}
\renewcommand{\theHequation}{supp.\arabic{equation}}
% Supplemental Material cover for the combined review PDF.
% 손으로 짠다 — 제목·저자·소속만, 초록 없이.
\thispagestyle{empty}
\vspace*{0.12\textheight}
\begin{center}
{\large\bfseries Supplemental Material\par}
\vspace{1.2em}
{\large Defect-limited thermal transport in AlN using pretrained machine-learning interatomic potentials\par}
\vspace{1.6em}
{Minseok Moon$^{a,1}$, Wonjun Choi$^{a,1}$, Seungwu Han$^{a,b,c,*}$, Youngho Kang$^{d,*}$\par}
\vspace{0.9em}
{\small\itshape $^{a}$Department of Materials Science and Engineering,
Seoul National University, Seoul, 08826, Republic of Korea\par}
{\small\itshape $^{b}$Research Institute of Advanced Materials,
Seoul National University, Seoul, 08826, Republic of Korea\par}
{\small\itshape $^{c}$Center for AI and Natural Sciences,
Korea Institute for Advanced Study, Seoul, 02455, Republic of Korea\par}
{\small\itshape $^{d}$Department of Materials Science and Engineering,
Incheon National University, Incheon 22012, Republic of Korea\par}
\vspace{1.2em}
{\small $^{*}$Corresponding authors.\par}
{\small $^{1}$These authors contributed equally to this work.\par}
\end{center}
\vfill
\clearpage

%======================================================================
% Supplementary figures and tables (formerly sections/supplement_body.tex)
%======================================================================
\begin{table}[htbp]
\caption{RTA and iterative three-phonon thermal conductivities of pristine AlN at 300~K.}
\label{smtab:solvercomparison}
\begin{tabular*}{\columnwidth}{@{\extracolsep{\fill}}lrrrrrr@{}}
\toprule
 & \multicolumn{2}{c}{RTA} & \multicolumn{2}{c}{Iterative} & \multicolumn{2}{c}{Change (\%)} \\
Model & in plane & $c$ axis & in plane & $c$ axis & in plane & $c$ axis \\
\midrule
DFT & 247.7 & 236.5 & 271.0 & 261.2 & 9.4 & 10.5 \\
omni-mpa & 235.2 & 207.8 & 259.1 & 231.2 & 10.2 & 11.3 \\
nano & 219.5 & 215.6 & 232.5 & 233.6 & 6.0 & 8.3 \\
ft-nano & 237.1 & 208.0 & 259.5 & 230.7 & 9.4 & 10.9 \\
\midrule
MACE-MH-1 & 236.8 & 225.0 & 257.0 & 245.3 & 8.5 & 9.0 \\
MACE-MPA-0 & 221.8 & 191.5 & 248.2 & 215.8 & 11.9 & 12.7 \\
\bottomrule
\end{tabular*}
\par\smallskip
{\footnotesize Conductivities are in W\,m$^{-1}$\,K$^{-1}$. RTA and iterative values come from the same calculation; percentage changes are relative to RTA.\par}

\end{table}

\begin{figure}[htbp]
\makebox[\textwidth][c]{\includegraphics{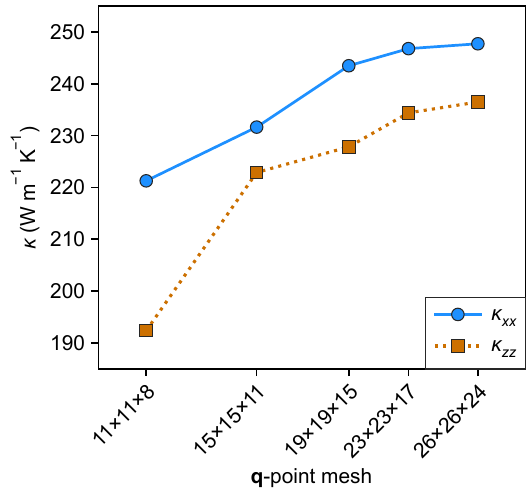}}
\caption{Convergence of the lattice thermal conductivity of pristine AlN with respect to the $\mathbf{q}$-point mesh, calculated in the RTA using the DFT force constants at 300~K, in plane ($\kL_{xx}$, solid line) and along the $c$ axis ($\kL_{zz}$, dotted line). The meshes are placed according to their total numbers of $\mathbf{q}$ points on a logarithmic scale. From the $23\times23\times17$ to the $26\times26\times24$ mesh, the conductivity changes by 0.4\% in plane and 0.9\% along the $c$ axis.}
\label{smfig:qmesh}
\end{figure}

% [not cited in the main text; commented out 2026-10-04]
%\begin{figure}[htbp]
%\makebox[\textwidth][c]{\includegraphics{figures/report/fig_cellsize_kappa_sm}}
%\caption{Thermal conductivity of AlN at 300~K obtained using SED defect rates from $6\times6\times4$, $6\times6\times8$, and $12\times12\times8$ supercells: (a) in plane and (b) along the $c$ axis. Filled points and dashed lines show iterative BTE results using SED and $T$-matrix defect rates, respectively, and open points denote pristine AlN.}
%\label{smfig:cellsizekappa}
%\end{figure}

\begin{table}[htbp]
\caption{Fine-tuning parameters for ft-nano.}
\label{smtab:ftproto}
\small
\setlength{\tabcolsep}{4pt}
\begin{tabular*}{\columnwidth}{@{\extracolsep{\fill}}p{0.34\textwidth}p{0.56\textwidth}@{}}
\toprule
Setting & Value \\
\midrule
\multicolumn{2}{@{}l}{\emph{Student architecture}} \\
\quad Cutoff radius & 5.0~\AA \\
\quad Interaction layers & 3 \\
\quad Channels & 32 \\
\quad Maximum angular momentum & $l_{\max}=2$ \\
\quad Initial weights & pretrained nano~\cite{oh2026nano} \\
\addlinespace
\multicolumn{2}{@{}l}{\emph{Teacher}} \\
\quad Model & SevenNet omni-mpa, multi-fidelity~\cite{kim2025} \\
\quad Fidelity channel & MPA \\
\quad Labels supplied & energy, force and stress from single-point calculations \\
\addlinespace
\multicolumn{2}{@{}l}{\emph{Optimization}} \\
\quad Loss & Huber, $\delta=0.01$ \\
\quad Loss coefficients & force 1.0, stress 0.01 \\
\quad Optimizer & Adam \\
\quad Learning rate & $2\times10^{-4}$, $\times0.985$ per epoch \\
\quad Batch size & 4 \\
\quad Epochs & 300 \\
\quad Adopted checkpoint & best validation error \\
\bottomrule
\end{tabular*}
\par\smallskip
{\footnotesize The student starts from nano weights and learns omni-mpa labels. Family-specific loss weights are listed in Table~\ref{smtab:ftdata}.\par}

\end{table}

\begin{table}[htbp]
\caption{Data groups and held-out errors before and after fine-tuning.}
\label{smtab:ftdata}
\small
\begin{tabular*}{\columnwidth}{@{\extracolsep{\fill}}p{0.36\textwidth}rrcc@{}}
\toprule
Group & $N_{\mathrm{atom}}$ & Frames & E RMSE (meV/atom) & F RMSE (meV/\AA) \\
\midrule
\multicolumn{5}{@{}l}{\emph{AlN bulk} ($w=1.0$)} \\
\quad Crystal MD, 300~K & 576 & 798 & $10.9\to0.01$ & $50\to3.9$ \\
\quad Crystal MD, 1100~K & 192--300 & 398 & $6.6\to0.09$ & $117\to16$ \\
\quad Melt--quench MD (amorphous) & 108 & 40 & $13.6\to0.06$ & $84\to11$ \\
\addlinespace
\multicolumn{5}{@{}l}{\emph{Charge-compensated oxygen defects}, $n\times[3\mathrm{O_N}{+}V_{\mathrm{Al}}]$, 300~K ($w=1.0$)} \\
\quad One complex, dispersed & 575 & 399 & $9.2\to0.41$ & $79\to9.3$ \\
\quad One complex, MC-annealed & 575 & 399 & $10.1\to0.13$ & $55\to6.0$ \\
\quad Two complexes, dispersed & 574 & 399 & $7.5\to0.83$ & $91\to11$ \\
\quad Two complexes, MC-annealed & 574 & 399 & $9.3\to0.22$ & $62\to7.7$ \\
\quad Dilute, large cells, rattled & 2151--4703 & 42 & $10.0\to0.18$ & $82\to16$ \\
\addlinespace
\multicolumn{5}{@{}l}{\emph{Amorphous Y--Al--O} (YAG composition; $w=3.0$)} \\
\quad Melt--quench--anneal MD & 120 & 330 & $12.5\to1.7$ & $215\to90$ \\
\quad MD of a quenched cell, 300--1200~K (aug.) & 120 & 44 & $8.0\to1.2$ & $188\to67$ \\
\addlinespace
\multicolumn{5}{@{}l}{\emph{Interfaces} ($w=4.0$)} \\
\quad AlN grain boundaries, relaxed/rattled & 864--1152 & 9 & $7.4\to0.27$ & $219\to31$ \\
\quad AlN grain-boundary MD, 300/900~K (aug.) & 864--1152 & 66 & $6.6\to0.87$ & $219\to60$ \\
\quad Y/O intergranular-film MD & 1152 & 44 & $4.9\to0.67$ & $302\to73$ \\
\quad AlN$|$film$|$AlN sandwich MD & 5688--5704 & 35 & $9.5\to1.4$ & $100\to55$ \\
\midrule
Total & & 3402 & $9.6\to0.69$ & $103\to28$ \\
\bottomrule
\end{tabular*}
\par\smallskip
{\footnotesize Arrows indicate errors before and after fine-tuning. $N_{\mathrm{atom}}$ is the cell size and $w$ is the family energy/force weight. Training uses 2,992 frames with a stratified 12\% validation split. ``Aug.'' denotes augmentation with 0.05~\AA\ rattling; MC-annealed cells contain bound complexes.\par}

\end{table}

\begin{figure*}[htbp]
\makebox[\textwidth][c]{\includegraphics{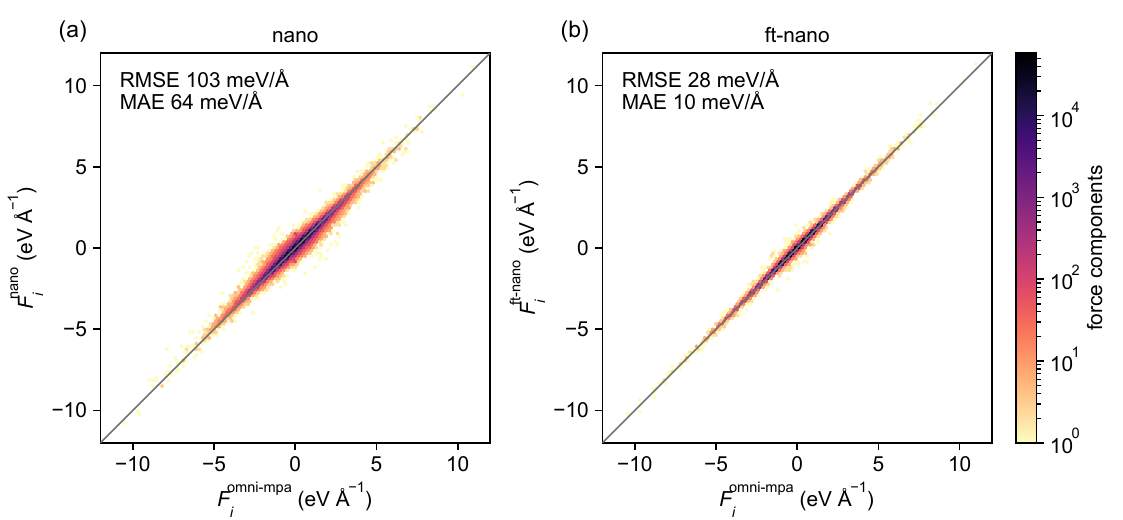}}
\caption{Force parity plots for (a) nano and (b) ft-nano against omni-mpa on the validation frames of Table~\ref{smtab:ftdata}.
 $F_i$ denotes a Cartesian component of the force on an atom, and superscripts indicate the potential. Colors indicate counts per bin, and diagonal lines indicate parity.}
\label{smfig:parity}
\end{figure*}

% ---- T-matrix supplement (formerly sections/tmatrix_supplement.tex) ----
%\begin{figure}[htbp]
%\makebox[\textwidth][c]{\includegraphics[width=190mm]{figures/report/fig_isolated_literature}}
%\caption{Isolated-defect scattering rates for (a) aluminum vacancies and (b) substitutional oxygen. $T$-matrix results of ft-nano are compared with charged-defect results of Li et al.~\cite{li2025aln}. Charge labels refer to the literature calculations.}
%\label{smfig:isolatedlit}
%\end{figure}

\begin{figure}[htbp]
\makebox[\textwidth][c]{\includegraphics{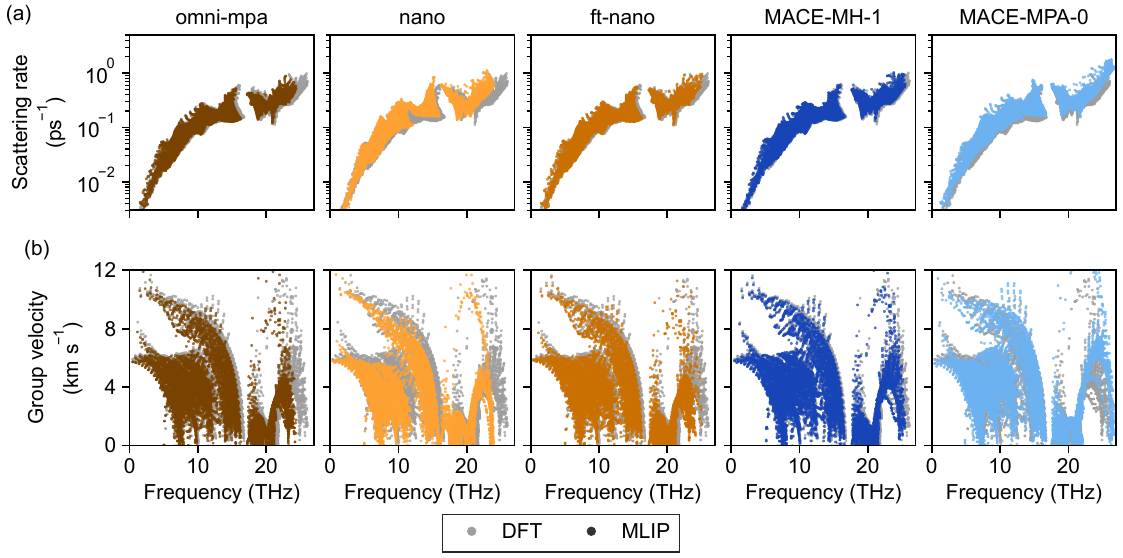}}
\caption{Phonon properties of pristine AlN calculated using DFT and MLIPs on a $26\times26\times24$ $\mathbf{q}$-point mesh: (a) three-phonon scattering rates at 300~K and (b) group velocities.}
\label{smfig:ratevg}
\end{figure}

% [not cited in the main text; commented out 2026-10-04]
%\begin{figure}[htbp]
%\makebox[\textwidth][c]{\includegraphics{figures/report/fig_kappa_spectral_ref}}
%\caption{Spectral lattice thermal conductivity of pristine AlN at 300~K from the iterative three-phonon BTE using DFT and MLIPs: (a) in plane and (b) along the $c$ axis.}
%\label{smfig:kappaspectral}
%\end{figure}

\begin{figure}[htbp]
\makebox[\textwidth][c]{\includegraphics{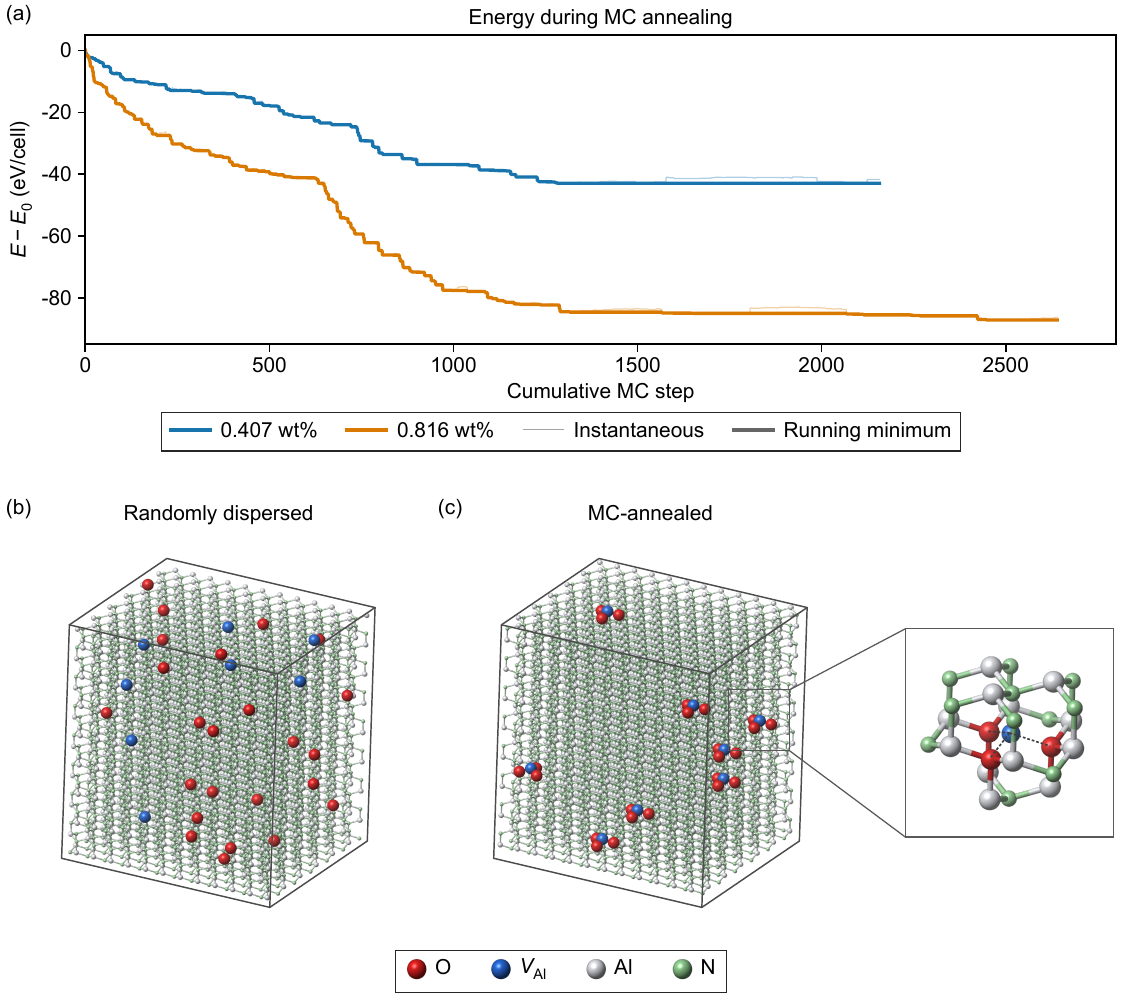}}
\caption{Monte Carlo sampling of oxygen-defective AlN. (a) Energy relative to the initial MC configuration from ft-nano at 1800~K for 0.407 and 0.816~wt\% oxygen in the 4608-site cell. Thin and thick lines denote instantaneous and running-minimum energies, respectively. (b, c) The 4608-site cell at 0.407~wt\% with (b) randomly dispersed and (c) MC-annealed defects; the inset shows a $V_{\mathrm{Al}}(\mathrm{O_N})_3$ complex and its nearest neighbors.}
\label{fig:mcanneal}
\end{figure}

\begin{figure}[htbp]
\makebox[\textwidth][c]{\includegraphics{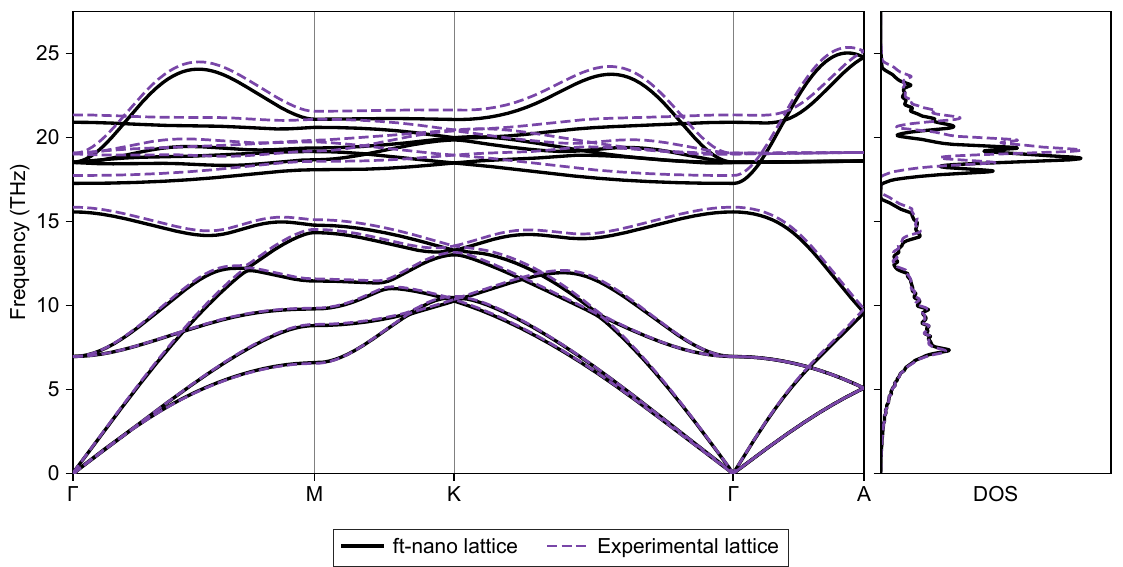}}
\caption{Harmonic phonon dispersions and densities of states of AlN calculated using ft-nano at the ft-nano and experimental lattice parameters of Table~\ref{smtab:latticebte}.}
\label{smfig:displattice}
\end{figure}

\begin{figure}[htbp]
\makebox[\textwidth][c]{\includegraphics{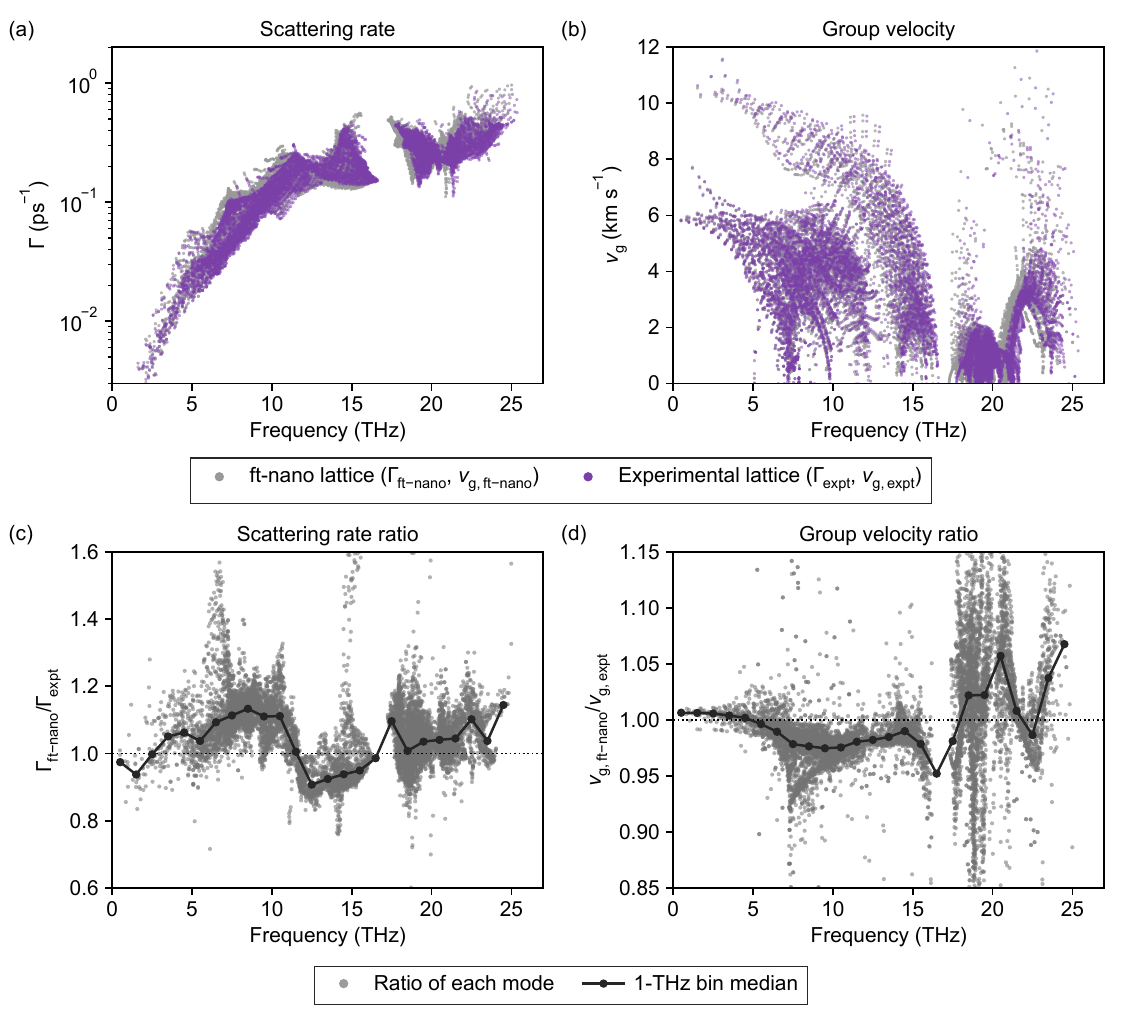}}
\caption{Three-phonon scattering rates and group velocities of pristine AlN at 300~K calculated using ft-nano at the ft-nano and experimental lattice parameters of Table~\ref{smtab:latticebte} on a $26\times26\times24$ $\mathbf{q}$-point mesh: (a) scattering rates $\Gamma$, (b) group velocities $v_{\mathrm{g}}$, and (c, d) their ratios, $\Gamma_{\mathrm{ft\mbox{-}nano}}/\Gamma_{\mathrm{expt}}$ and $v_{\mathrm{g,ft\mbox{-}nano}}/v_{\mathrm{g,expt}}$, for each phonon mode. Modes are paired by wave vector and branch index; lines connect the medians over 1-THz frequency bins, and dotted lines denote unity.}
\label{smfig:ratevglattice}
\end{figure}

\begin{table}[htbp]
\caption{Thermal conductivities of pristine AlN calculated using ft-nano at two sets of lattice parameters. The cell is fixed at the listed lattice parameters and only the atomic positions are relaxed. Conductivities are obtained from the iterative three-phonon BTE at 300~K and are given in W\,m$^{-1}$\,K$^{-1}$.}
\label{smtab:latticebte}
\small
\begin{tabular*}{\columnwidth}{@{\extracolsep{\fill}}lcccc@{}}
\toprule
Lattice & $a$ (\AA) & $c$ (\AA) & $\kL_{xx}$ & $\kL_{zz}$ \\
\midrule
ft-nano & 3.1282 & 5.0119 & 259.5 & 230.7 \\
Expt.$^{a}$ & 3.113 & 4.982 & 282.0 & 257.9 \\
\bottomrule
\end{tabular*}
\par\smallskip
{\footnotesize The second- and third-order force constants are recomputed on each lattice. $^{a}$Rounded room-temperature values, within 0.064\% of those in Table~\ref{tab:reference}.\par}
\end{table}

% [not cited in the main text; commented out 2026-10-04]
%\begin{figure}[htbp]
%\makebox[\textwidth][c]{\includegraphics{figures/report/fig_tmatrix_models_sm}}
%\caption{Harmonic $T$-matrix scattering rates of the $V_{\mathrm{Al}}(\mathrm{O_N})_3$ complex calculated using (a) DFT, (b) nano, and (c) ft-nano. The DFT rates are obtained from a $4\times4\times3$ supercell (192 atoms) at the same $c_{\mathrm{def}}$.}
%\label{smfig:tmatrixmodels}
%\end{figure}

\begin{figure}[htbp]
\makebox[\textwidth][c]{\includegraphics{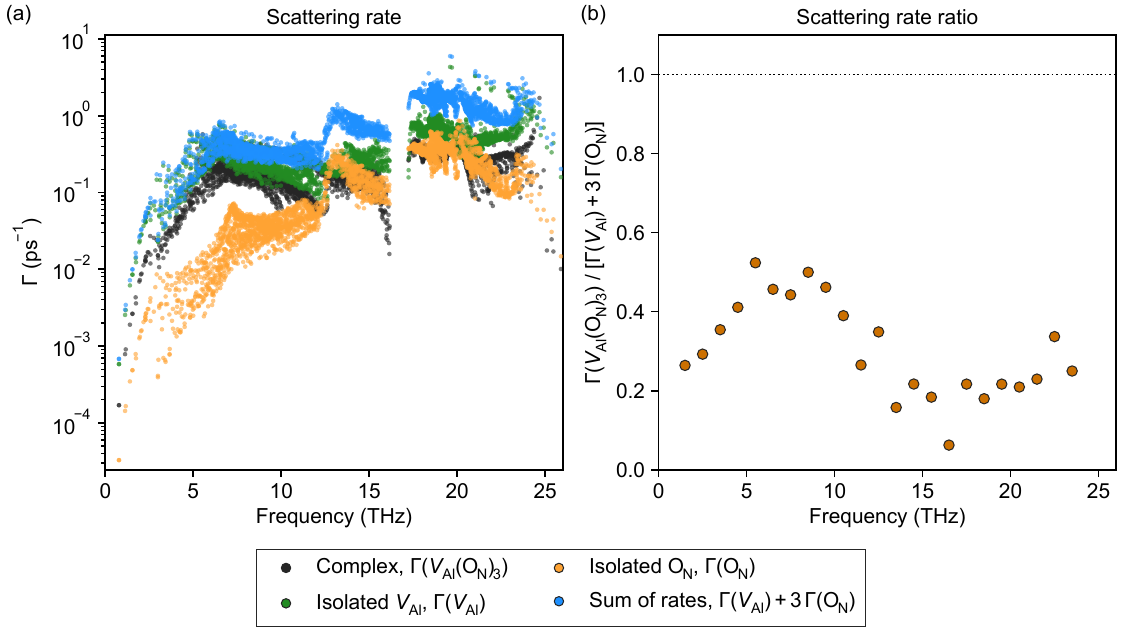}}
\caption{Harmonic $T$-matrix scattering rates of the $V_{\mathrm{Al}}(\mathrm{O_N})_3$ complex and of its constituent defects calculated using ft-nano. (a) Scattering rates of the complex, an isolated $V_{\mathrm{Al}}$, and an isolated $\mathrm{O_N}$, and the sum of the isolated rates, $\Gamma(V_{\mathrm{Al}})+3\,\Gamma(\mathrm{O_N})$, for each phonon mode. Each isolated defect is calculated in its own supercell, so the sum contains no interaction between the vacancy and the oxygen atoms. All rates correspond to the same number of complexes, or of each constituent defect, per primitive cell (0.407~wt\% oxygen). (b) Ratio of the 1-THz bin medians of the complex rate to those of the summed rate; the dotted line denotes unity.}
\label{smfig:cvp}
\end{figure}

\begin{figure}[htbp]
\makebox[\textwidth][c]{\includegraphics{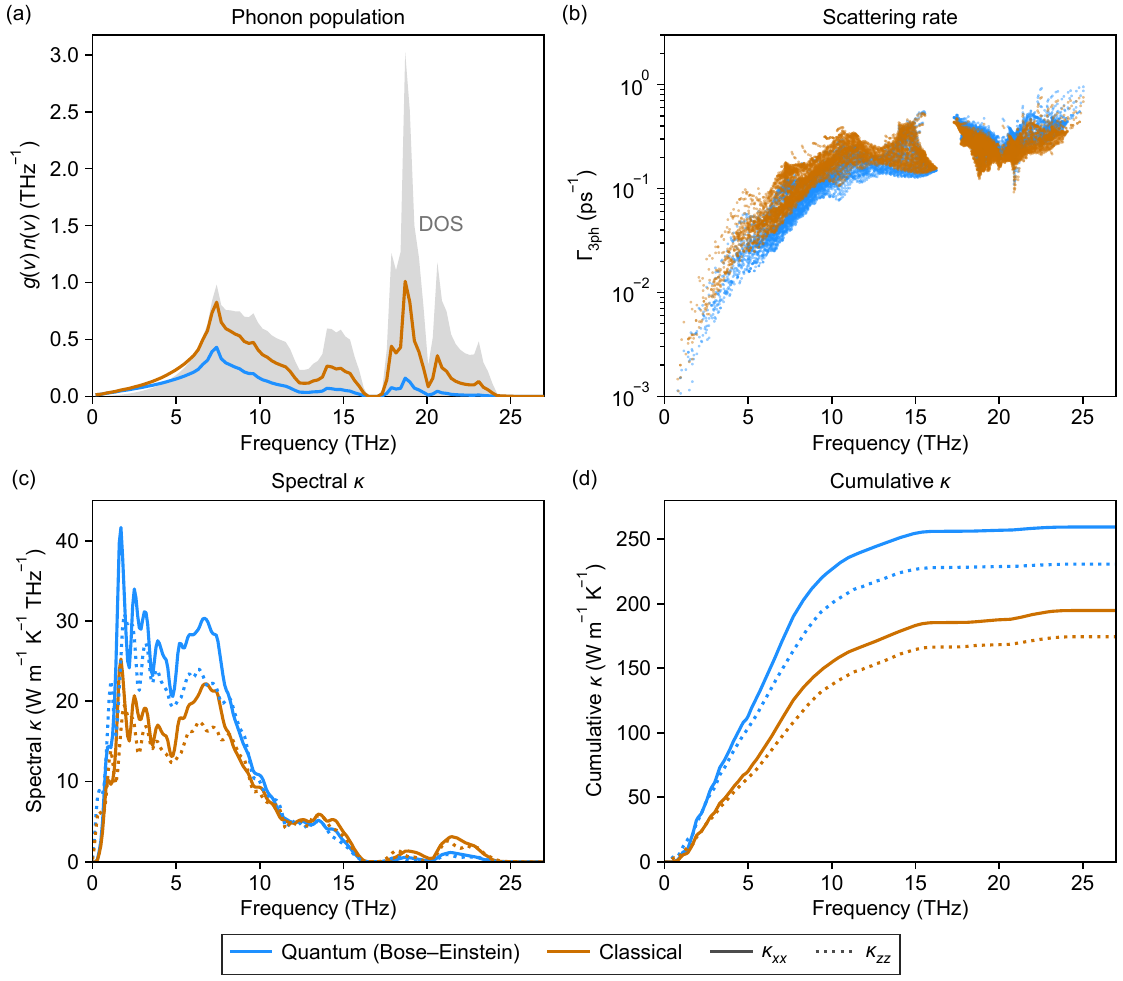}}
\caption{Effect of phonon statistics on the lattice thermal conductivity of pristine AlN calculated using ft-nano at 300~K on a $26\times26\times24$ $\mathbf{q}$-point mesh. (a) Phonon population $g(\nu)n(\nu)$ with Bose--Einstein and classical ($k_{\mathrm{B}}T/h\nu$) occupations; the shaded area is the phonon density of states $g(\nu)$. (b) Three-phonon scattering rates of each mode. (c) Spectral and (d) cumulative thermal conductivities from the iterative BTE in plane ($\kL_{xx}$, solid lines) and along the $c$ axis ($\kL_{zz}$, dotted lines). The classical calculation uses classical occupations in the three-phonon scattering rates and the classical heat capacity, $k_{\mathrm{B}}$ per mode. The classical conductivities, 194.8 and 174.4~W\,m$^{-1}$\,K$^{-1}$, are 25\% lower than the quantum values, 259.5 and 230.7~W\,m$^{-1}$\,K$^{-1}$.}
\label{smfig:classical}
\end{figure}

%\begin{figure}[htbp]
%\makebox[\textwidth][c]{\includegraphics{figures/report/fig_kappa_lattice_sm}}
%\caption{Thermal conductivity of AlN as a function of oxygen concentration at 300~K calculated using ft-nano at the ft-nano, experimental, and LDA lattice parameters of Table~\ref{smtab:latticebte}: (a) in plane and (b) along the $c$ axis. Lines show iterative BTE results using $T$-matrix defect rates recomputed on each lattice, and open points denote pristine AlN. The green line shows Slack's experimental fit~\cite{slack1973}.}
%\label{smfig:kappalattice}
%\end{figure}

% References only when this file is compiled on its own (main.tex already prints them).
\ifdefined\inmaindocument\else
\clearpage
\bibliographystyle{elsarticle-num}
\bibliography{refs}
\fi

\end{document}